\documentclass[aps,pra,twocolumn,superscriptaddress,nofootinbib]{revtex4-1}
\usepackage{amssymb,graphicx,color}
\usepackage[intlimits]{amsmath}
\usepackage[english]{babel}
\usepackage[colorlinks]{hyperref}
\usepackage{comment}
\def\beq{\begin{equation}}
\def\eeq{\end{equation}}
\def\bea{\begin{eqnarray}}
\def\eea{\end{eqnarray}}
\def\barr{\begin{array}}
\def\earr{\end{array}}

\begin{document}

%\title{Entanglement in an interacting double quantum dot system: The curious role of degeneracy}

\title{Quantum entanglement and transport in a non-equilibrium interacting double-dot system: The curious role of degeneracy}

\author{Anirban Dey}
\affiliation{Indian Institute of Science Education and Research Bhopal, India}
\author{Devendra Singh Bhakuni}
\affiliation{Indian Institute of Science Education and Research Bhopal, India}
\author{Bijay Kumar Agarwalla}
\affiliation{Department of Physics, Indian Institute of Science Education and Research, Pune 411008, India}
\author{Auditya Sharma}
\email{auditya@iiserb.ac.in}
\affiliation{Indian Institute of Science Education and Research Bhopal, India}

\date{\today}

\begin{abstract}
We study quantum entanglement and its relation to transport in a
non-equilibrium interacting double dot system connected to electronic
baths. The dynamical properties in the non-interacting regime are
studied using an exact numerical approach whereas the steady state
properties are obtained following the well-known non-equilibrium
Green's function (NEGF) approach. By means of mutual information and
concurrence we explore the connection between the quantum correlations
in the system and the current flowing through the dots.  It is
observed that entanglement between the dots is heavily influenced by
the degeneracy or the lack thereof, of the dot levels. In the
non-degenerate case, the concurrence falls sharply when the applied
bias crosses a certain critical value.  In contrast when the dot
energy levels are degenerate, the concurrence reaches a very high
asymptotic value of $1/2$.  When interactions are switched on, the
degeneracy is lifted, and once again concurrence falls to zero beyond
a critical value of the applied bias. Lastly it is observed that the
concurrence can be made to reach almost the value of $1.0$ if the
chemical potential in both baths are made very large (while keeping
the sign the same) provided the dot levels are kept degenerate within
the non-interacting limit. A combination of NEGF method, brute-force
numerics and asymptotics are employed to corroborate our findings.
\end{abstract}

%\pacs{}

\maketitle

\section{Introduction}
Although an enormous body of knowledge has been created within the field of
mesoscopic systems~\cite{beenakker1991theory,
  meir1991transport,rokhinson1999kondo,matveev1996coulomb,sztenkiel2007electron,
  you1999electron,zimbo2008elec}, an understanding of the role of
quantum correlations in these systems is rather primitive.  Recent
evidence~\cite{auditya2015landauer,bosonicdot} suggests that transport
across quantum dots is intimately connected to the quantum
correlations that develop between various subsystems. The question of
how general these connections are, is still a matter of
investigation. Non-equilibrium properties of interacting systems are
notoriously hard to study, and there are few  precious results on
quantum correlations in such systems. Previous
work~\cite{auditya2015landauer,bosonicdot} pertains only to the
non-interacting limit, and moreover only mutual
information~\citep{mutualinfo1,mutualinfo2,mutualinfo3,amico2008entanglement} has been
studied, which contains not only quantum but also classical
correlations.  An understanding of the relationship between purely
quantum correlations and transport, remains a pressing open question,
even within the non-interacting limit. The other pressing question is
the role of interactions.  The present work is an attempt to make
progress along both of these directions.

Here, we show how
concurrence~\citep{wootters1998entanglement,cho2017quantum,deng2004fermionic,
  zanardi2002fermionic,nehra2018many,wu2011quant}, which is an
excellent measure of entanglement in fully mixed states may be studied
in the presence of interactions. Furthermore our work pertains to the
non-equilibrium regime, where we attempt a comparison of the transport
properties and the quantum correlations that develop in the
system. While the full time-evolution is accessible in the
non-interacting limit, we concentrate on the steady state properties
of the interacting model. Our system consists of a spinless quantum
double dot connected to left and right baths with inter-dot
interactions. In the current work we have
  theoretically studied the entanglement properties of a nonequilibrium interacting double dot system. A
  natural division of subsystems for the purpose of entanglement is
  the two dots, and we compute concurrence between the dots and investigate
  how it is connected to the current flowing through the dots.

  Entanglement is very hard to measure experimentally. Quantum
  mechanics tells us that the expectation values of observables have
  the form $Tr(\rho O)$, where $O$ is the operator corresponding to
  the relevant observable. However, von Neumann entropy is given by
  $-Tr(\rho\log(\rho))$. $\log(\rho)$ is not an observable, and in
  fact, is a quantity that is dependent on the state itself. Therefore
  there are very subtle conceptual difficulties associated with
  whether it is even meaningful to measure von Neumann entropy. Hence,
  \emph{indirect} access to entanglement is of great interest. In this
  context, there have been works that have tried to access
  entanglement via other quantities like current~\citep{auditya2015landauer,bosonicdot}, conductance~\cite{PhysRevLett.120.146801}, and
  quantum noise~\cite{klich2009quantum,auditya2015landauer}. A lot of work has been done previously to understand
  the transport through the interacting quantum dot sytems and
  phenomena like the Coulomb blockade
  effect~\citep{Averin1986,waugh1995single} and Kondo
  effect~\citep{kondo1964resistance} have been understood both
  theoretically and
  experimentally~\citep{Cronenwett540,ding2019competition,goldhaber1998kondo,brotons2019coulomb}. Furthermore,
  within the Kondo regime, the entanglement between the magnetic
  impurity and the baths has been studied via different measures of
  entanglement~\citep{PhysRevLett.120.146801,Chung2018,PhysRevB.97.155123,PhysRevB.95.115106,PhysRevB.96.075157}.
  The indirect detection of entanglement via the conductance has also
  been proposed~\citep{PhysRevLett.120.146801}. Clever strategies for a frontal attack on this problem continue
  to be of great current importance~\cite{wei2019direct}.
\begin{figure}
\includegraphics[scale = 1]{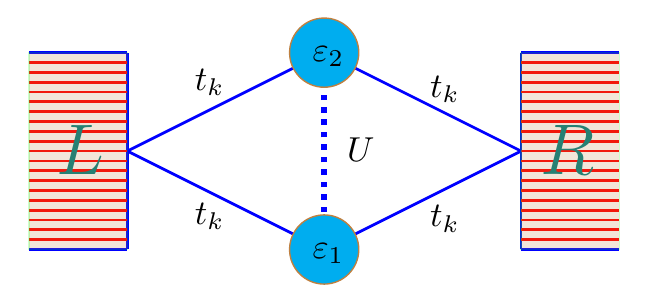}
\caption{(Color online) A schematic of the parallel double quantum dot setup under study. Here $\epsilon_1$ and $\epsilon_2$ represents the energies of the dots. $U$ is the Coulomb interaction strength. $t_k$ is the tunneling amplitude between the dots and the baths ($L$ and $R$) and is assumed to be identical here.}
\end{figure}

The non-equilibrium Green's Function (NEGF)
method~\citep{schwinger1961brown,MeirCurrent,haug2008quantum,RLandauer1970Quantumtransport,
  Wang2014,jauho1994time,rammer1986quant} has been one of the most
powerful analytical tools to tackle non-equilibrium physics. It has
been used widely and successfully in the domain of quantum dots to
study steady state transport and mesoscopic properties, and we too
adopt this method. This technique is also useful in the
non-interacting limit where we have access to the properties of the
system in a deeper way. Here we take recourse to the equation of
motion (EOM)
technique~\citep{levy2013steady,van2014noneq,Yuan2012noneq} to
calculate the Green's Function for both non-interacting and
interacting models. For the non-interacting model we have followed
both NEGF and exact numerical techniques in order to test for their
mutual agreement, thus strengthening the quality of our
results. For the interacting model we have used the
  Hartree-Fock approximation for decoupling the higher order
  correlators in such a way that the interaction effect on one site
  depends upon the average electron number of the other site. This is
  a simple mean field way of considering the interaction and yet has been argued to be
  legitimate for a large range of the Coulomb interaction
  strength~\citep{meir1991transport,sun2002double,levy2013steady,dhar2006nonequilibrium,
    agarwal2006conductance,lamba2000transport,you1999electron}. There
  are also some advanced methods such as numerical renormalization
  group (NRG), real time renormalization group (RTRG), functional
  renormalization group (fRG), hierarchical quantum master equation (HQME) etc.~\citep{linder2019renorm,PhysRevLett.120.146801,erpenbeck2018extending,hartle2013decoherence,zheng2009numerical,
jin2008exact,okamoto2016hierarchical,PhysRevB.90.245426}, but they are rather complicated and could be useful in extracting
finer structure.

The study of quantum correlations within the non-interacting limit is
facilitated by special tricks which exploit Wick's Theorem, even for non-equilibrium systems~\cite{auditya2015landauer}.
%which is inherent in the density matrix of such systems, even within the
%nonequilibrium regime~\cite{auditya2015landauer}. 
This allows for a reduction of a $2^{L}\times 2^{L}$ reduced density
matrix problem to an $L\times L$ correlation matrix problem~
\cite{peschel2003RDM} with $L$ being the number of sites in the
subsystem of interest, thus providing access to large system sizes.
Exploiting this trick, it is possible to investigate mutual
information between the quantum dots and the baths.  However in the
presence of interactions, the correlation matrix approach no longer
holds, making the computation of quantities that
involves reduced density matrix of the baths, a notoriously hard problem. 

In the present work, we
focus on the entanglement (mutual information, concurrence) that
develops between two quantum dots when they are driven out-of-equilibrium,
and investigate how it is correlated with the current flowing through
the dots.
The key findings of our paper are as follows:  In the non-interacting limit, current and mutual information between the dots and the bath, have similar behaviour as in the single dot
problem~\cite{auditya2015landauer}. The physics is strongly dependent
on whether or not the two dot energy levels are degenerate. In the
presence of degeneracy, both concurrence and mutual information between the
dots, have finite steady state values, in contrast to their zero
steady state values in the absence of degeneracy. The non-interacting
degenerate system allows for the possibility of generating very high
entanglement between the dots. By cranking up the left and right leads
to a large and identical chemical potential, it becomes possible to
make the two dots to reach unit concurrence value.  Turning on
interactions results in a lifting of the degeneracy leading to the
concurrence again attaining a zero value in the steady state.

The plan of the paper is as follows. We begin with an introduction of
the model Hamiltonian in Sec.~II. In Sec.~III we first discuss the
non-interacting limit of the model following the NEGF approach and an
exact numerical technique and present analytical and numerical results
for charge transport, mutual information and concurrence. In Sec.~IV
we extend our study to the interacting Hamiltonian. Finally in Sec.~V we summarize
our findings.

\section{Model Hamiltonian} \label{Model hamiltonian}
We consider an open quantum system with two parallel quantum dots connected to both left
and right baths. The baths consist of non-interacting spinless fermions. 
In the non-interacting limit, the coherent tunnelling term between the
dots is absent whereas in the interacting limit the dots exchange
energy via Coulomb interaction. The total Hamiltonian can be separated
into different components:
\begin{eqnarray} \label{eq1}
%\begin{split}
H &=& H_{D} + H_{L} + H_{R} + H_{LD} + H_{RD}, \nonumber \\ 
\end{eqnarray}
where $H_D$ represents the dot Hamiltonian 
\begin{equation}
H_{D} = \epsilon_{1}\, n_{1} + \epsilon_{2} \, n_{2}+ U n_{1}n_{2}.
\end{equation}
Here $n_i=d_i^{\dagger}d_i, i=1,2$ represents the number operator with
$d_{i}^{\dagger}$ $(d_{i})$ being the electronic creation
(annihilation) operator for the $i$-th dot with energy
$\epsilon_{i}$. $U$ denotes the Coulomb interaction strength.
\begin{equation}
H_{\alpha} =\sum_{k} \epsilon_{k\alpha} n_{k\alpha}, \quad \quad \alpha =L, R,
\end{equation}
represents the bath Hamiltonian modelled as an infinite collection of
non-interacting fermions with momentum index $k$ and corresponding
energy $\epsilon_k$. $n_{\alpha k} = c^{\dagger}_{k \alpha} c_{k
  \alpha}$ is the number operator with $c_{k\alpha}^{\dagger}
(c_{k\alpha})$ being the creation (annihilation) operators for
$k^{\text{th}}$ mode.  Finally, the system-bath coupling Hamiltonian
$H_{\alpha D}$ is given as
\begin{equation}
H_{\alpha D} = \sum_{i=1,2}^{}t_{k\alpha}^{i} c_{k \alpha}^{\dagger}d_{{i}}+{\rm h.c.} , \quad \alpha=L,R.
\end{equation}
Here $t^i_{k\alpha}$ represents the tunneling amplitude between the $i^{\text{th}}$ dot and the $k^{\text{th}}$ mode of bath. 
This coupling information is encoded in the spectral density of the baths as follows:
\begin{equation} \label{eq2}
\Gamma_{\alpha}^{ij}(\epsilon)=2\pi\sum_{k} t^i_{k\alpha} (t^j_{k\alpha})^{*} \delta(\epsilon - \epsilon_{k\alpha}).
\end{equation}
Throughout the paper we set $\hbar=e=k_{B}=1$.
The dots are driven out-of-equilibrium by maintaining the electronic
baths at different chemical potentials $\mu_{\alpha}$. In the present work, we keep the two bath temperatures identical $T_{L} = T_{R}$, although 
in principle this is another knob that can yield a non-equilibrium scenario.
We then follow the dynamical as well as steady state properites of various thermodynamic
and quantum informatic observables. In what follows, we will first
consider the non-interacting limit ($U=0$) and present analytical and
numerical results followed by results for the interacting case ($U
\neq 0$).

\section{Non-interacting dots ($U=0$)}
We begin with the NEGF formalism and discuss briefly the method
to compute various two-point correlation functions (Green's functions)
for the subsystem as well as for the baths. In terms of these Green's
functions different thermodynamic observables can be computed.
\subsection{NEGF Formalism}
NEGF is a powerful tool to study transport properties of out-of-equilibrium many-body quantum systems. 
Within the NEGF method, there are several well established approaches
to compute Green's functions, like the Keldysh diagrammatic
method~\cite{agarwalla2016tuna, keith2009quant}, and the
equation-of-motion (EOM)
technique~\cite{niu1999equ,levy2013steady,Galperin2007inel}. Here we
adopt the EOM approach to compute the Green's functions.
The first task is to define the time-ordered Green's function for the dots:
\begin{equation}
G_{0,ij}(t, t') = -i  \langle T \, d_{i}(t)d_{j}^{\dagger}(t')\rangle,\quad i=1,2.
\label{GFs}
\end{equation} 
Here the time dependence in the operators represents the Heisenberg
picture with operators evolving with the full Hamilonian $H$. $T$
represents the time-ordered operator and the average $\langle ... \rangle$ is computed with respect to the initial
  condition. Following the Heisenberg's EOM, the EOM for the Green's
functions can be obtained.  A cascade of such equations of motion can then be written down,
introducing new Green's functions at every next step, until a closure
among the Green's functions is attained to terminate the procedure.
For non-interacting systems, such a closure is possible to obtain which
then yields exact analytical results.  In contrast, for
interacting systems, the closure is attained by employing different
approximation schemes.
The EOM for Eq.~\ref{GFs} is given as
\begin{equation}
\Big[i \frac{\partial}{\partial t} - \epsilon_i \Big]G_{0, ij}(t,t') = \delta_{ij} \delta(t-t')  + \sum_{k\alpha=L,R} (t^i_{k, \alpha})^{*} G^{0}_{kj}(t,t'),
\label{EOM-GFs}
\end{equation}
where 
\begin{equation}
G^{0}_{kj}(t,t') = -i\langle T c_{k \alpha}(t) d_j^{\dagger}(t') \rangle
\end{equation}
is the time-ordered version of a mixed Green's function involving the
dot and the lead. In a similar manner, a differential equation for
this mixed Green's function can be obtained:
\begin{equation}
\Big[i \frac{\partial}{\partial t} - \epsilon_{k\alpha} \Big] G^{0}_{kj}(t,t') = \sum_{l=1,2} t^{l}_{k\alpha} G^{0}_{lj}(t,t').
\end{equation}
Introducing the bare Green's functions for the subsystem (dots) and the leads:
\begin{eqnarray}
\Big[i \frac{\partial}{\partial t} - \epsilon_i \Big]g_{ij}(t,t') &=& \delta_{ij} \delta(t-t') \nonumber \\
\Big[i \frac{\partial}{\partial t} - \epsilon_{k \alpha} \Big]g^{\alpha}_{k}(t,t') &=& \delta(t-t'), 
\end{eqnarray} 
and defining the self-energy for the electronic baths
\begin{equation}
\Sigma^{\alpha}_{ij}(t-t') = \sum_{k} t_{k\alpha}^i g_{k \alpha}(t-t') t_{k \alpha}^{j}, \alpha=L,R, 
\end{equation}
we can write down a formal solution for Eq.~\ref{EOM-GFs} as 
\begin{equation}
{\bf G}_0(t,t')= {\bf g}(t,t') + \int^{t} dt_1 \int^{t} dt_2 \, {\bf g}(t,t_1) {\bf \Sigma}(t_1, t_2) {\bf G}_0(t_2,t').
\label{sol-GF}
\end{equation} 
Here ${\bf \Sigma}={\bf \Sigma}_L + {\bf \Sigma}_R$ is the total
self-energy, additive in both left and right leads, and is associated
with the transfer of electrons between the leads and the dots. The
bold notation refers to a matrix in the subsystem space.
In a general non-equilibrium setup, a similar equation as in Eq.~\ref{sol-GF} can
be obtained for the contour-ordered Green's function where instead of
real times $t$ and $t'$ one introduces contour time variables $\tau$ and
$\tau'$ that runs on a complex time plane~\citep{haug2008quantum}. We therefore write, for
non-equilibrium systems,
\begin{equation}
{\bf G}_0(\tau, \tau')= {\bf g}(\tau,\tau') + \int \! d\tau_1 \! \int d\tau_2  {\bf g}(\tau,\tau_1) {\bf \Sigma}(\tau_1, \tau_2) {\bf G}_0(\tau_2,\tau').
\end{equation} 
From this contour-ordered Green's function, applying Langreth theorem~\citep{haug2008quantum}, one then has access to all other real time
Green's functions namely, retarded $({\bf G}^r_0(t,t'))$, advanced
$({\bf G}^a_0(t,t'))$, lesser $({\bf G}^<_0(t,t'))$ and greater $({\bf
  G}^>_0(t,t'))$ components.
In the steady state limit, the expression for these Green's functions
can be simplified by taking advantage of the time-translational
symmetry. In this limit, the Green's functions depend only on the
relative time difference $|t - t'|$ and performing a Fourier
transformation one obtains for retarded and advanced components
\begin{equation}
{\bf G}_0^{r/a}(\epsilon)= {\bf g}^{r/a}(\epsilon) + {\bf g}^{r/a}(\epsilon) {\bf \Sigma}^{r/a}(\epsilon) {\bf G}_0^{r/a}(\epsilon)
\end{equation}
which can be rewritten as 
\begin{equation}
{\bf G}_0^{r/a}(\epsilon) = \Big[\epsilon {\bf I} - {\bf H}_D -{\bf \Sigma}^{r/a}(\epsilon) \Big]^{-1}
\label{retarded-advanced}
\end{equation}
with $H_D = {\rm diag} (\epsilon_1, \epsilon_2)$.
The lesser and greater components follow the Keldysh equation 
\begin{equation}
{\bf G}_0^{</>}(\epsilon)= {\bf G}_0^{r}(\epsilon) {\bf \Sigma}^{</>}(\epsilon) {\bf G}_0^{a}(\epsilon), 
\label{keldysh}
\end{equation}
where ${\bf \Sigma}^{r,a,<,>}= {\bf \Sigma}_L^{r,a,<,>}+ {\bf
  \Sigma}_R^{r,a,<,>}$ are different components of the total
self-energy given by:
\begin{eqnarray}\label{self}
&&{\bf \Sigma}_{\alpha}^{r/a}(\epsilon)  = \mp i\frac{\Gamma_{\alpha}(\epsilon)}{2} \begin{pmatrix} 1 & 1 \\ 1 & 1\end{pmatrix}, \nonumber  \\ &&
{\bf \Sigma_{\alpha}^{<}} (\epsilon) = i f_{\alpha}(\epsilon) \Gamma_{\alpha}(\epsilon)\begin{pmatrix}1 & 1 \\1 & 1 \end{pmatrix}  \quad\text{and}\nonumber\\ 
&&{\bf \Sigma_{\alpha}^{>}} (\epsilon) =- i (1-f_{\alpha}(\epsilon)) \Gamma_{\alpha}(\epsilon)\begin{pmatrix}1 & 1 \\1 & 1\end{pmatrix}.
\end{eqnarray}
In writing the expressions for ${\bf \Sigma}_{\alpha}^{r/a}(\epsilon)$ we
have ignored the real part which is responsible for the
re-normalization of the bare dot energies. We further assume identical
coupling between the bath and the dots $\Gamma^{ij}_{\alpha} =
\Gamma_{\alpha}$. Note that the non-trivial temperature and bias
information is contained in the lesser and greater components of the
Green's functions, given in Eq.~\ref{keldysh}.

%%%%%%%%%%%%%%%%%%%%%%%%%%%%%%%%%%%%%%%%%%%%%%%%%%%%%%%%%%%%%%%%
%%%%%%%%%%%%%%%%%%%%%%%%%%%%%%%%%%%%%%%%%%%%%%%%%%%%%%%%%%%%%%%%
\subsection{Exact Numerical Approach and Dynamical Protocol}
\begin{figure*}
\includegraphics[scale=0.75]{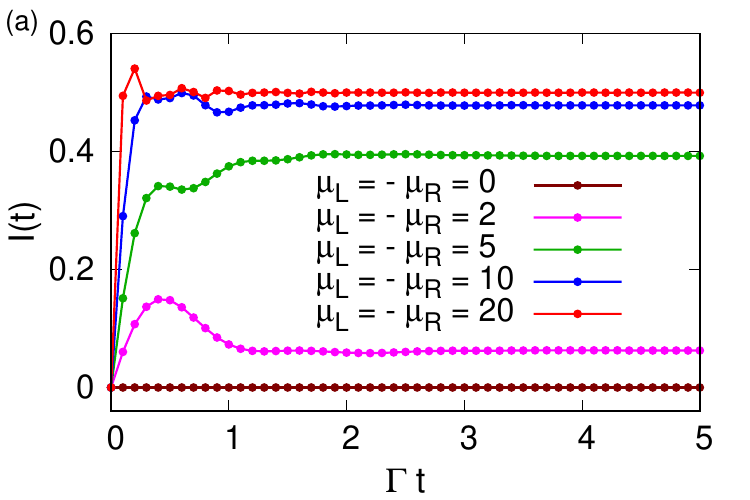}
\includegraphics[scale=0.75]{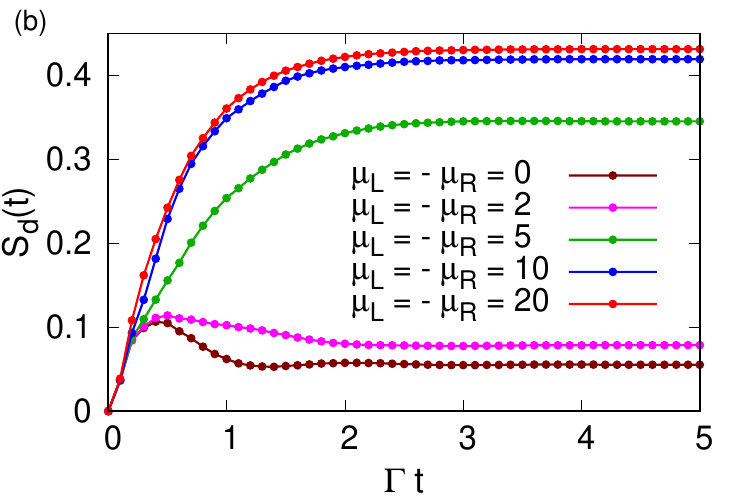}
\includegraphics[scale=0.75]{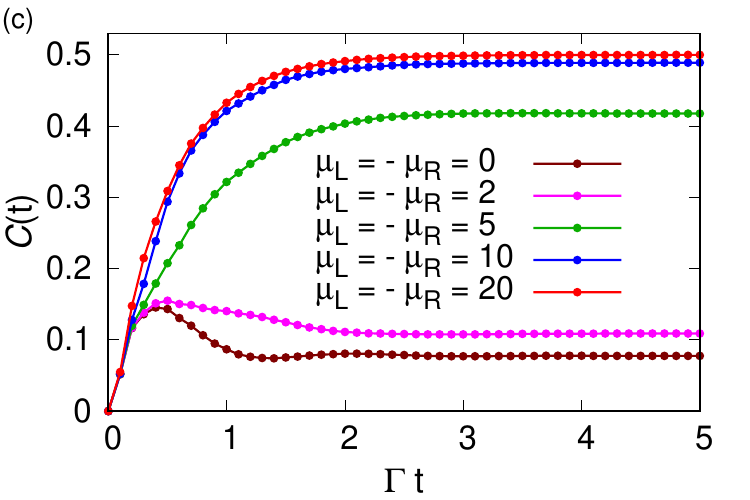}
\caption{(Color online)  Quantum dynamics of (a) charge current flowing through the dots, (b) mutual information, and (c) concurrence between the two dots for the degenerate case $(\epsilon_{1} = \epsilon_{2} = 3.5)$. The other parameters are: $N_L=N_R=128, \epsilon_c=20, k_{B} T=0.05$.}
\label{curr}
\end{figure*}
\begin{figure*}
\includegraphics[scale=0.75]{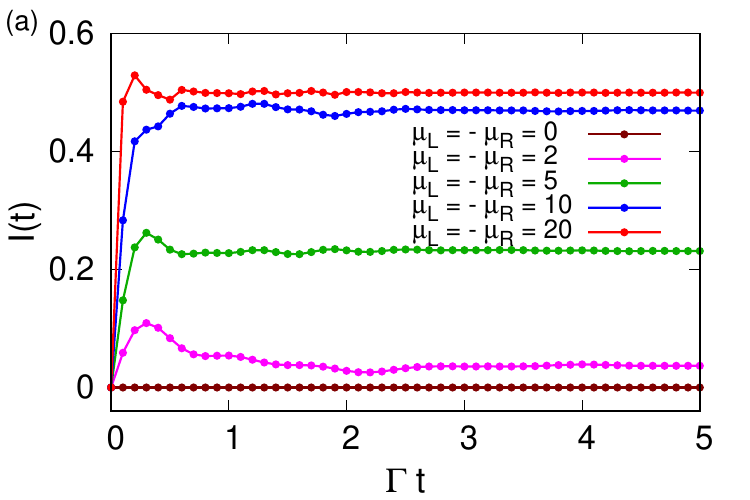}
\includegraphics[scale=0.75]{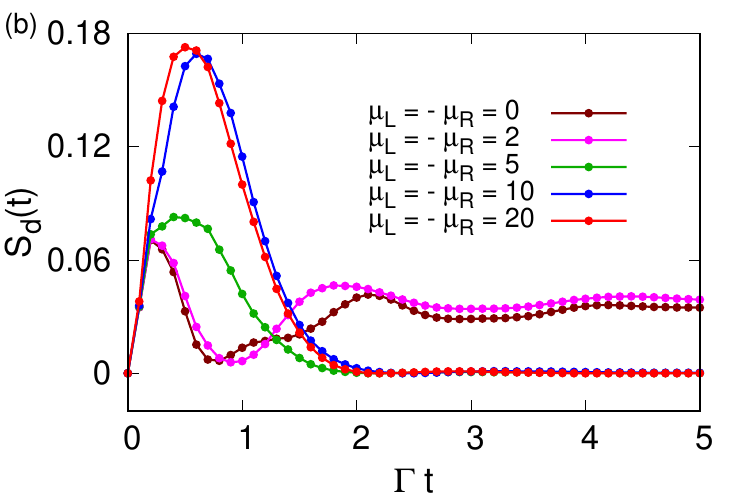}
\includegraphics[scale=0.75]{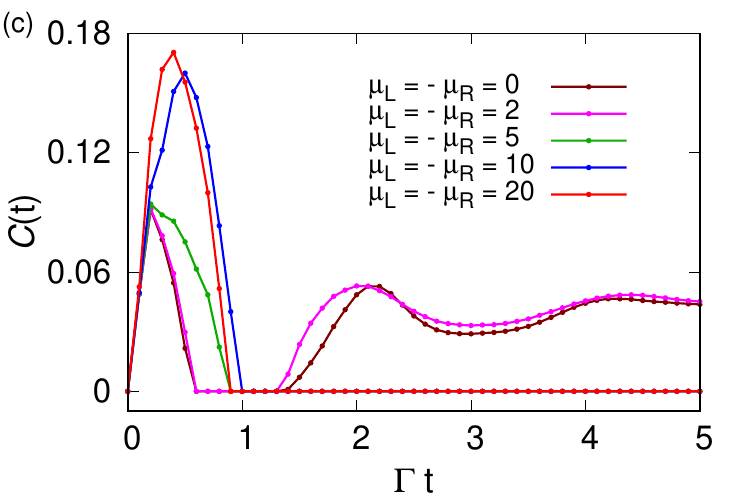}
\caption{(Color online) Quantum dynamics of (a) charge current flowing through the dots, (b) mutual information, and (c) concurrence between the two dots for the non-degenerate case $(\epsilon_{1} = 3.5,\ \epsilon_{2} = 6.5)$. The other parameters are the same as in Fig.~\ref{curr}.}
\label{ent2}
\end{figure*}
We next consider an exact numerical approach~\citep{auditya2015landauer} to simulate the non-interacting double quantum dot setup. This method gives direct access to the density matrix of the entire system (dots, baths). In this approach, we discretize the bath spectrum and 
consider a large but equispaced finite number of levels. The coupling between the baths and the dots are then fixed by integrating the spectral function (Eq.~\ref{eq2}) in a small energy window $\Delta\epsilon$ around $\epsilon_{k}$ and further by imposing the wide-band limit we obtain 
\begin{equation}
t_{k\alpha}=\sqrt{\frac{\Gamma_{\alpha}(\epsilon_k)\Delta\epsilon}{2\pi}} \quad = \quad \sqrt{\frac{\Gamma_{\alpha}\Delta\epsilon}{2\pi}}.
\end{equation}
%At $t=0$, the couplings between the system and baths are turned on and the time dynamics is governed by the full Hamiltonian $H$ as given in Eq.~\ref{eq1} with $U=0$. 
The final discretized form of the Hamiltonian $H$ can then be written as   
\begin{equation}\label{eq:19a}
H = \sum_{k=1}^{N} \zeta_{k} c_{k}^{\dagger}c_{k} + \sum_{k=1}^{N-2} t_{k}(c_{k}^{\dagger}c_{N-1} + c_{N-1}^{\dagger}c_{k})+t_{k}(c_{k}^{\dagger}c_{N} + c_{N}^{\dagger}c_{k}),
\end{equation}
where $N$ is the total number of levels including the baths and the
dots, $N = N_{L}+N_{R}+2$ with $N_{L}$ and $N_{R}$ being the energy-levels
in the left and right baths, respectively. The indices $N-1$ and $N$
are used for the two dots with $c_{N-1}^{\dagger}(c_{N-1})=d_{1}^{\dagger}(d_{1}), c_{N}^{\dagger}(c_{N})=d_{2}^{\dagger} (d_{2})$, and the  corresponding energies are, $\zeta_{N-1}=\epsilon_{1}$ and $\zeta_{N}=\epsilon_{2}$. 
The explicit form of the Hamiltonian in the matrix form can be expressed as  
\begin{equation} \label{eq: 19}
H= 
  \begin{pmatrix}
  
    \zeta_{1} & 0 & \hdots & 0 & 0 & 0 & \hdots & \hdots& t_{1} & t_{1}\\
     0 &\zeta_{2} & 0 & \hdots & \hdots &\hdots & \hdots & 0 & t_{2} & t_{2}\\
     \vdots &\vdots & \ddots & \vdots & \vdots & \vdots & \vdots & \vdots & \vdots & \vdots\\
    ... & ... & \hdots & \zeta_{N_L} & 0 & ... & ... & ... & t_{N_{L}}& t_{N_{L}}\\
     0 & 0 & \hdots & 0 & \zeta_{1} & 0 & \hdots & \hdots & t_{1} & t_{1}\\
     0 & 0 & \hdots & 0 & 0 & \zeta_{2} & \hdots & \hdots & t_{2} & t_{2}\\
    \vdots & \vdots & \vdots & \vdots & \vdots & \vdots & \ddots & \vdots & \vdots & \vdots\\
     0 & 0 & \hdots & 0 & 0 & 0 & \hdots & \zeta_{N_{R}} & t_{N_{R}} & t_{N_{R}}\\
     t_{1} & t_{2} & \hdots & t_{N_{L}} & t_{1}& t_{2}& \hdots & t_{N_{R}} & \epsilon_{1} & 0\\
     t_{1} & t_{2} & \hdots & t_{N_{L}} & t_{1}& t_{2}& \hdots & t_{N_{R}} & 0 & \epsilon_{2}
  \end{pmatrix}.
\end{equation}  
Note that, one can write the above Hamiltonian $H$ in a diagonal form by introducing new set of fermionic operators $a_{\beta}$, defined as $a_{\beta} =
\sum_{i=1}^{N} \psi_{\beta}(i)c_{i}$, where $\psi_{\beta}(i)$ are coefficients of the eigenvectors of the Hamiltonian coupling matrix. 

Next we choose a decoupled initial condition for the global density matrix, written as a tensor product of the density matrix of each part as 
\begin{equation}
\rho(0) = \rho_{L}(0)\otimes\rho_{R}(0)\otimes\rho_{D}(0),
\end{equation}
where the left and the right fermionic baths are distributed according to the  grand canonical distribution
\begin{align}
\rho_{L}(0) = \frac{\exp(-\beta(H_{L}-\mu_{L}\sum_{k\in L} c_{k}^{\dagger} c_{k}))}{Z_{L}},\nonumber\\ \rho_{R}(0) = \frac{\exp(-\beta(H_{R}-\mu_{R}\sum_{k\in R} c_{k}^{\dagger} c_{k}))}{Z_{R}}.
\end{align}
Here both left and right baths are maintained at a fixed inverse temperature $\beta= 1/T$ and at different chemical potentials $\mu_L$ and $\mu_R$ respectively.
The density matrix for the dots is represented as :
\begin{align}
\rho_{D}(0) = \rho_{D_1} \otimes \rho_{D_2} \\
\rho_{D_1}(0) = n_{0}d_{1}^{\dagger}d_{1} + (1 - n_{0})d_{1}d_{1}^{\dagger}\\
\rho_{D_2}(0) = n_{0}d_{2}^{\dagger}d_{2} + (1 - n_{0})d_{2}d_{2}^{\dagger},
\label{initialcond}
\end{align} 
where $n_{0}$ is the initial population of the dots. Given this initial setup, the 
global density matrix at any arbitrary time is then simulated following the unitary time evolution $\rho(t) = e^{- i H t}\rho(0)e^{i H t}$.

For our numerical simulations, we have considered the baths with
$N_{L/R}=128$ levels and finite cut-off for the energy spectrum $-\epsilon_c \leq \epsilon \leq \epsilon_c$ with $\epsilon_c=20$. We have fixed the step size as $\Delta \epsilon = 2\epsilon_{c}/({N_L}-1)$. These parameters provide convergence upto the desired accuracy. 
Using this scheme we provide results for the current, mutual information and concurrence.

Note that, in the non-interacting case, the Gaussian nature of the initial state allows the use of Wick's theorem to translate the problem of calculating the density matrix to a simpler problem of calculating the correlation matrix: $C_{ij}=\langle c_{i}^{\dagger}c_{j}\rangle$~\citep{auditya2015landauer,dhar2012nonequilibrium}. Different physical observables can then be computed following this correlation matrix approach.

\begin{figure*}[t]
\includegraphics[scale=0.75]{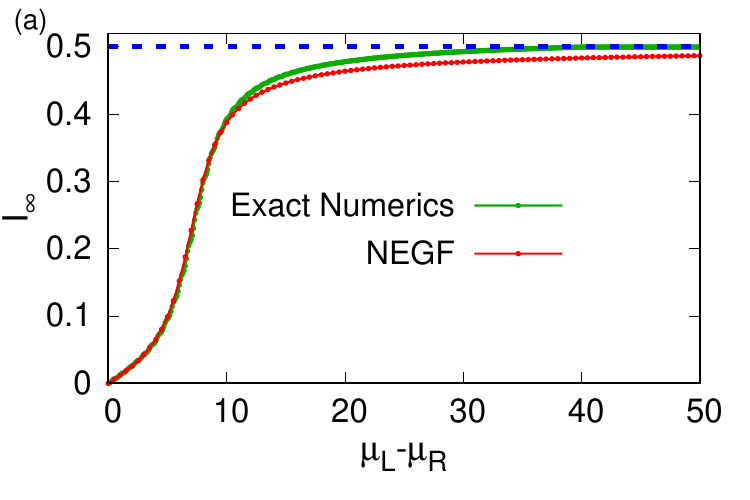}
\includegraphics[scale=0.75]{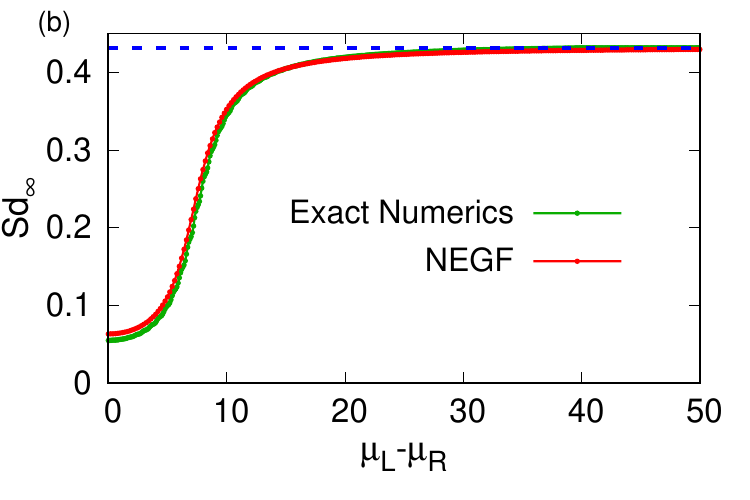}
\includegraphics[scale=0.75]{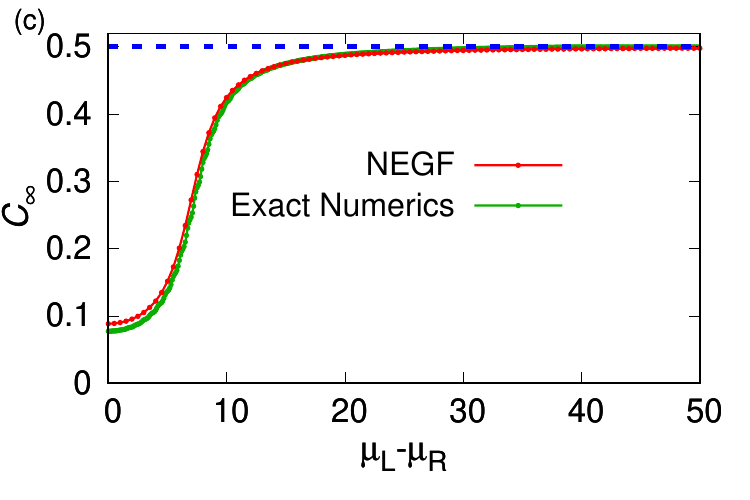}
\caption{(Color online) Steady state properties: a comparison of results from exact numerics and NEGF approach as a function of applied bias voltage for (a) charge current (b) mutual information and (c) concurrence between the dots for the same dot energies: $(\epsilon_{1} = 3.5, \epsilon_{2} = 3.5)$. The other parameters are: $N_L=N_R=128, \epsilon_c=20, k_{B} T=0.05$.   The blue dotted line in (a) and (c) corresponds to an asymptotic value of $\frac{1}{2}$ whereas in (b) it corresponds to a value $0.43145$. }\label{stadys1}
\end{figure*}
\begin{figure*}[t]
\includegraphics[scale=0.75]{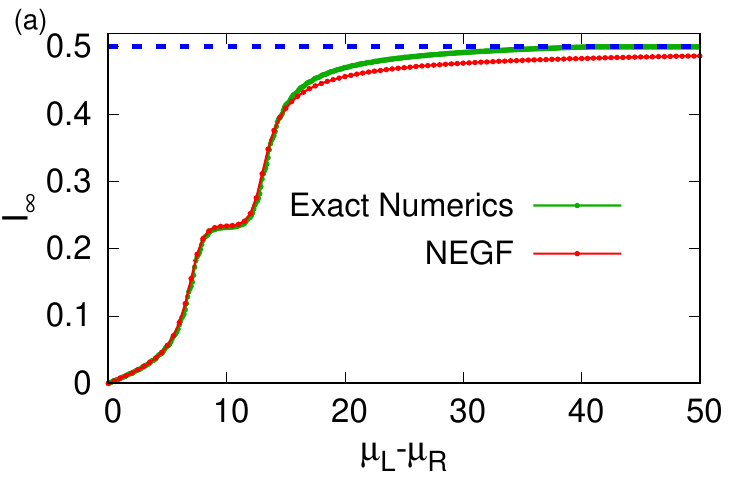}
\includegraphics[scale=0.75]{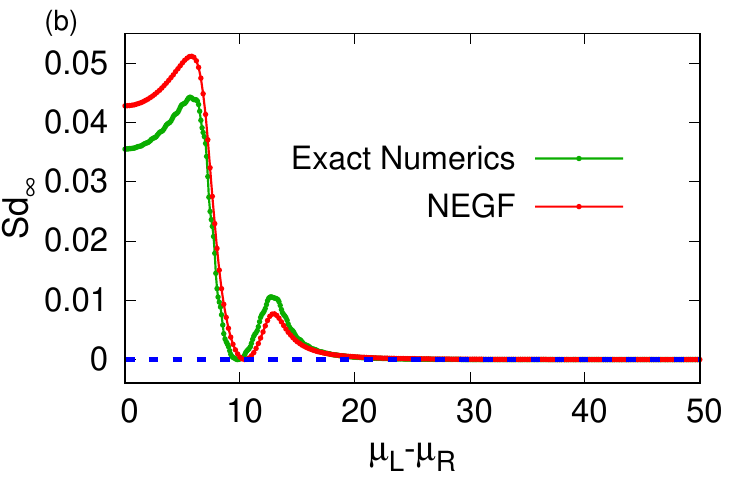}
\includegraphics[scale=0.75]{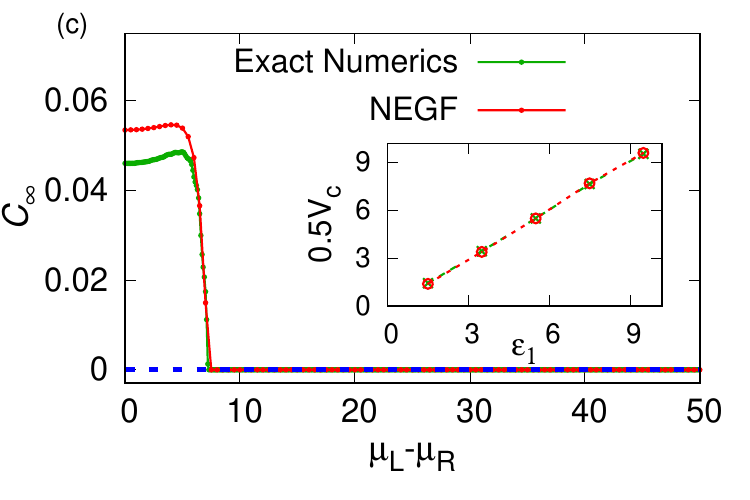}
\caption{(Color online) Steady state properties: a comparison of results from exact numerics and NEGF approach as a function of applied bias voltage for (a) charge current (b) mutual information and (c) concurrence between the dots for different dot energies: $(\epsilon_{1} = 3.5, \epsilon_{2} = 6.5)$. The inset in (c) is a plot for the critical value of voltage $(V_c = 2\epsilon_1)$ where the concurrence drops to zero as a function of lower energy level $\epsilon_1$. The other parameters are the same as in Fig.~(\ref{curr}).
The blue dotted line in (b) and (c) corresponds to a value of $0$ whereas in (a) it corresponds to a value $\frac{1}{2}$. }\label{stadys2}
\end{figure*}

\subsection{Results for charge current, mutual information and concurrence} 
We will begin by defining three relevant observables for this setup, namely the charge current, mutual information and concurrence.

{\it Charge current:} 
We first investigate the charge current flowing out of one of the leads. The current is defined as the rate of change of occupation number in a particular bath 
\begin{eqnarray}\label{eq21}
I_{L/R}(t) = -  \Big\langle\frac{d N_{L/R} }{dt}\Big\rangle ,
\label{cur-gen}
\end{eqnarray}
with $N_{L/R}= \sum_{k} c^{\dagger}_{k L/R} c_{k L/R}$. A formal expression for the steady state charge current for an arbitrarily interacting quantum junction can be written down in terms of the Green's functions and is given by the celebrated Meir-Wingreen formula ~\citep{haug2008quantum,meir1992landauer}
\begin{equation}
I_{L/R} =  \int_{-\infty}^{\infty}\, \frac{d\epsilon}{2 \pi} \, {\rm Tr} \Big[ {\bf G}^{<}(\epsilon) {\bf \Sigma}_{L/R}^{>}(\epsilon) -  {\bf G}^{>}(\epsilon) {\bf \Sigma}_{L/R}^{<}(\epsilon)\Big].
\end{equation}
Here ${\bf G}^{</>}(\epsilon)$ refers to the Green's function for an interacting subsystem and ${\bf \Sigma}_{L/R}^{</>}(\epsilon) $ is the usual self-energy term involving the leads. 
Upon symmetrization, $I = (I_{L} - I_{R})/2$, the expression for current simplifies to
\begin{eqnarray}
I &=&  \frac{i}{2} \int _{-\infty}^{\infty}\, \frac{d\epsilon}{2 \pi} \, {\rm Tr} \Big[\big({\bf \Gamma}_L(\epsilon) - {\bf \Gamma}_R(\epsilon)\big) \,  {\bf G}^{<}(\epsilon)   \\
&+& \Big(f_L(\epsilon)\, {\bf \Gamma}_L(\epsilon) - f_R(\epsilon)\, {\bf \Gamma}_R(\epsilon) \Big) \big({\bf G}^r(\epsilon)-{\bf G}^a(\epsilon)\big) \Big]\nonumber. 
\end{eqnarray}
For a symmetric junction i.e., ${\bf \Gamma}_{L}(\epsilon)={\bf \Gamma}_{R}(\epsilon)={\bf \Gamma}(\epsilon)/2$  the first term of the above expression vanishes and one obtains a formula for steady state current for a general interacting system as
\begin{equation}\label{eq29}
I = \frac{1}{4 } \int _{-\infty}^{\infty}\, \frac{d\epsilon}{2 \pi} \, {\rm Tr} \Big[ {\bf \Gamma}(\epsilon) {\bf A}(\epsilon) \Big] \big(f_L(\epsilon)-f_R(\epsilon)\big) \Big],
\end{equation}
where ${\bf A}(\epsilon) = i \big({\bf G}^r(\epsilon) - {\bf
  G}^a(\epsilon)\big)$ is defined as the spectral function matrix for
the subsystem.

We now present the numerical results for the charge current. We display the steady state results obtained following the NEGF method whereas for the charge current dynamics we follow the exact numerical approach. Note that, for the numerical approach, we compute the following expression for current which can be obtained using  Eq.~\ref{cur-gen} and the Heisenberg equation of motion,
\begin{eqnarray}\label{eq22}
I_{L/R}(t) = i\sum_{k \in L(R)}t_{k} \big\langle(c_{k}(t)c_{N-1}^{\dagger}(t) - c_{k}^{\dagger}(t)c_{N-1}(t)) \big \rangle \nonumber \\ + t_{k} \big\langle(c_{k}(t)c_{N}^{\dagger}(t) - c_{k}^{\dagger}(t)c_{N}(t)) \big \rangle.\quad 
\end{eqnarray}
These correlators can be calculated for arbitrary time following the correlation matrix approach.   

In Fig.~(\ref{curr}a) and (\ref{ent2}a) the dynamics of the charge current is shown both for degenerate and non-degenerate cases. After a short transient dynamics, in each case, the current saturates to a steady state value and this saturation value depends on the bias difference applied across the dots.  However, interestingly, in the large bias limit, the asymptotic value is indifferent for both degenerate and non-degenerate dots and saturates to a value $1/2$.

We next compare the steady state current as a function of the bias voltage in Fig.~(\ref{stadys1}a) and Fig.~(\ref{stadys2}a). A nice agreement between the exact numerical approach and NEGF results is obtained for both degenerate and non-degenerate dots. The slight mismatch occurs due to the finite discretization of the bath spectrum. Note that, in the non-degenerate case, one observes an additional plateau in the current as compared to the degenerate case. Due to the symmetric choice of the chemical potential i.e.,  $\mu_L = -\mu_R$, the plateau starts to appear at $\mu_L-\mu_R\approx 2 \epsilon_1 $ providing a single channel for the electron to flow
and disappears when $\mu_L-\mu_R\approx 2 \epsilon_2$ 

{\it Mutual information:} To understand the corresponding effects in the context of correlations in
the system, we study the mutual information~\citep{mutualinfo1,mutualinfo2,mutualinfo3,amico2008entanglement} between the dots and
the baths as well as between the two dots. The mutual information
for a bipartite system $A$ and $B$ is defined as
\begin{equation} \label{eq23}
S = S_{A} + S_{B} - S_{AB} ,
\end{equation} 
where $S_{A} = -\text{Tr}_A(\rho_{A}\mbox{ln}\rho_{A})$, $S_{B} = -\text{Tr}_B(\rho_{B}\mbox{ln}\rho_{B})$ and $S_{AB} = -\text{Tr}_{A+B}(\rho_{AB} \mbox{ln}\rho_{AB})$ corresponds to the von Neumann entropies of $A,\ B$ and the composite system $A+B$, respectively. For example, to compute the mutual information between the two dots, $A$ and $B$ should correspond to individual dots.

Typically, for a general interacting system, the computation of the bath Von Neumann entropy is a challenging problem. However, in the
non-interacting limit, following
the correlation matrix
approach~\citep{peschel2003RDM,peschel2012special}, the von Neumann
entropy $S_{G}$ of a subspace $G$ can be computed as and is given
as~\citep{auditya2015landauer}
\begin{equation} \label{eq: 3.16}
S_{G} = \sum_{\sigma = 1}^{N_{G}}[-(1 - C_{\sigma})\mbox{ln}(1 - C_{\sigma}) - C_{\sigma}\mbox{ln}(C_{\sigma})],
\end{equation}
where $C_{\sigma}$ are the eigenvalues of the correlation matrix defined within subspace $G$ and $N_{G}$ are the total number of sites in that subspace. For example, for subsystem consisting of two quantum dots $N_G=2$ and $C_{ij}={\rm Tr}\Big[\rho(t) d_i^{\dagger} d_j\Big], i,j=1,2$ following the notations used in Eq.~\ref{eq1} with $U=0$.

In Fig.~(\ref{curr}b) and Fig.~(\ref{ent2}b) we display the dynamics of mutual information $s_d(t)$ between the two dots for both degenerate and non-degenerate cases. As can be seen, similar to the charge current, after a transient regime a steady state is achieved for the mutual information. It starts with a zero value due to the choice of our initial condition (see Eq.~\ref{initialcond}). For the same dot energies and in the large bias limit, the mutual information saturates to a finite value $\approx 0.43145$ whereas it vanishes for different dot energies.  This dynamical behaviour is very similar also for the concurrence between the dots, as shown in Fig.~(\ref{curr}c) and Fig.~(\ref{ent2}c) and a detailed discussion of this follows ahead.  The steady state results for the mutual information are displayed in Fig.~(\ref{stadys1}b) and
Fig.~(\ref{stadys2}b). The signature of the plateau in the current for the non-degenerate case (Fig.~(\ref{stadys2}a)) gets reflected in the mutual information by a peak.

\begin{figure}
\includegraphics[scale=1.0]{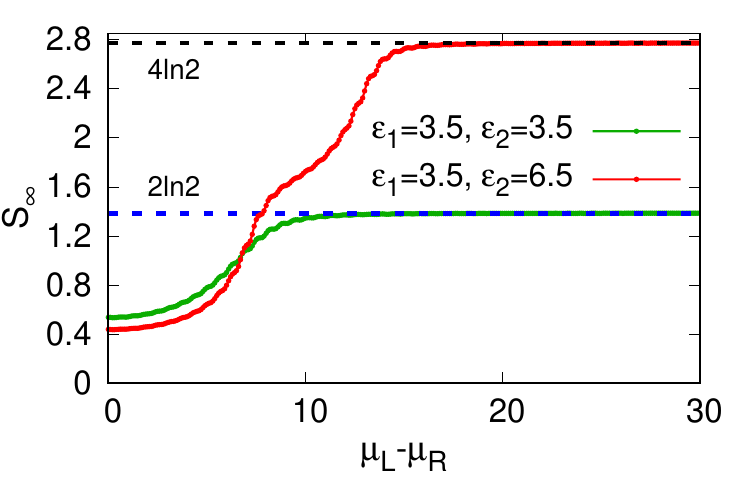}
\caption{(Color online) The steady state value of mutual information between the dots and baths as a function of bais voltage . The other parameters are: $N_L=N_R=128, \epsilon_c=20, k_{B} T=0.05$. The black dotted line corresponds to an asymptotic value of $4\text{ln}2$ whereas the blue dotted line corresponds to an asymptotic value of $2 \text{ln}2$. }\label{stadys3}
\end{figure}

We also investigate the steady state mutual information between the dots and the bath as a function of the bias voltage in (Fig.~\ref{stadys3}). Interestingly, in this case,  the qualitative behaviour is the same as that of the current flowing through
the dots. In the non-degenerate case a plateau appears followed by a
saturation value  $4\,\text{ln}2$, whereas in the degenerate case the mutual information
directly saturates to a value $2\, \text{ln}2$. These saturation values for various observables in the asymptotic limit can be obtained analytically and is discussed in the subsection D. 

{\it Concurrence:}  Although the exact numerical method allows to calculate the mutual
information between the dot system and the baths in the non-interacting regime, a generalization to the interacting case is not trivial. However the double-dot system lends itself naturally to study two-site entanglement between the dots.
A particularly useful quantity to measure this type of entanglement is called concurrence ~\citep{wootters1998entanglement,cho2017quantum}.
For a number conserving Hamiltonian the reduced density matrix for two sites can be written as \citep{nehra2018many}
\begin{equation}\label{eq32}
\rho_{12} =\begin{bmatrix}
&u &0 &0 &0\\
&0 &w_{1} &z^{*} &0\\
&0 &z &w_{2} &0\\
&0 &0 &0 &v\\
\end{bmatrix},
\end{equation}
where, $u=\langle(1-n_1)(1-n_2)\rangle,   w_{1}=\langle(1-n_1)n_2\rangle, w_{2}=\langle n_1(1-n_2)\rangle, v=\langle n_{1}n_{2}\rangle \   \text{and}\  z=\langle d_{2}^{\dagger}d_{1}\rangle$. These various correlators can be calculated using Wick's theorem. 

The concurrence is then given by 
\begin{equation}\label{eq33}
\mathcal{C} = 2 \, \text{max}(0,|z|-\sqrt{uv}).
\end{equation}

In Fig.~(\ref{stadys1}c) and Fig.~(\ref{stadys2}c) we display the steady state concurrence between the dots as a function of bias voltage. In the non-degenerate limit (Fig.~(\ref{stadys2}c)), at low bias, concurrence starts with a very small value and suddenly falls off sharply to zero. This sharp fall occurs when $\mu_L-\mu_R \approx 2 \epsilon_1$. This transition value of the bias voltage is plotted against the lower dot energy level $\epsilon_1$ (inset in Fig.~(\ref{stadys2}c)), setting $\epsilon_2=10$. A linear dependence is obtained following both NEGF and exact numerics. For small bias, neither of the dot channels
are activated and therefore the two dots can be present in a mixed entangled state. In
this regime, an increasing bias voltage results in an increasing
entanglement. Further increase in the bias voltage $\mu_L-\mu_R \geq 2 \epsilon_1$ allows electron to tunnel through
by populating the lowest energy level of the dot and keeping the higher energy dot 
empty. This leads to a separable state for the dots and concurrence drops to zero. This trend continues even when $\mu_L-\mu_R \geq 2 \epsilon_2$ as the density matrix for the two dots always remain separable.

In contrast, for the degenerate limit, the concurrence value increases
monotonically with applied bias Fig.~(\ref{stadys1}c) before reaching
an asymptotic steady state value of $1/2$ in the large bias limit.
Because of degeneracy among the dot levels, an incoming electron from
the bath can not differentiate between the two levels leading to a
formation of a mixed entangled state. It further remains as an
entangled state for higher bias voltage.

\begin{figure}
\includegraphics[scale=1.0]{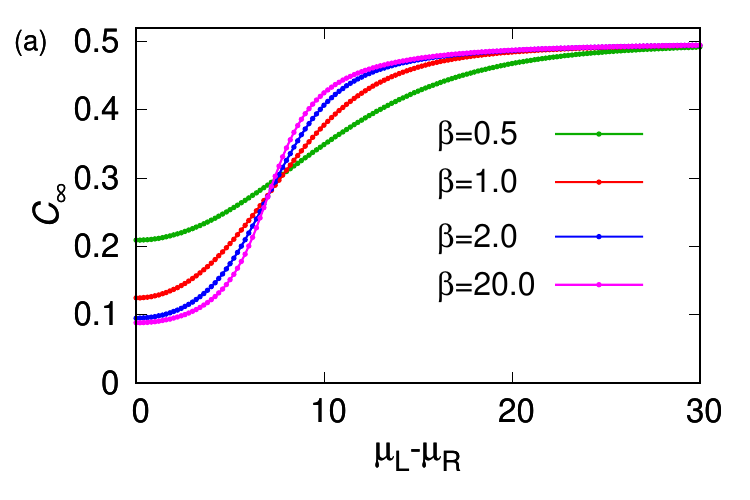}
\includegraphics[scale=1.0]{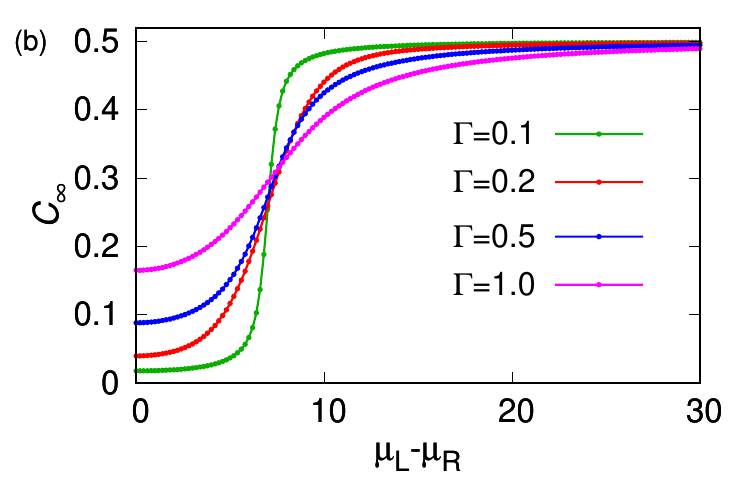}
\caption{(Color online) The (a) temperature and (b) coupling
  dependence of concurrence as a function of bias voltage using the
  NEGF approach. Increasing the temperature or the coupling leads to
  an increase in the concurrence for small bias voltage. The
  saturation value on the other hand, remains independent of both temperature and coupling strength. The dot
  energies are: $\epsilon_1 = \epsilon_2 = 3.5$ and $\Gamma = 0.5$ for
  (a) while $k_B T =0.05$ for (b).}
\label{betagammadep}
\end{figure}

The other factors which can influence the concurrence are the
temperature ($T$) and the coupling between the bath and the dots
($\Gamma$) as studied in Fig.~\ref{betagammadep}. On varying the temperature, the concurrence also changes
because of the thermal transport which initiates the transport of
electrons even before the bias reaches the critical value. This also
reflects in the concurrence as it can be seen that a finite correlation
between the electrons of the two sites exists even for very low bias. The same argument holds on varying the coupling $\Gamma$. A
greater value of the coupling means the transport is active even for small bias voltage
and a large finite value of concurrence is obtained. It is also worth noting that the saturation value of concurrence is independent
of both temperature and coupling.

\subsection{Analytical results in the asymptotic limit}
In this subsection we provide asymptotic large bias limit results for the charge current, mutual information, and concurrence.

We begin with the two point correlators $\langle d_{i}^{\dagger}d_{j}\rangle$ which are related to lesser or greater version of the corresponding Green's function and can be computed by performing an integral in the energy domain
\begin{equation}
-i {\bf G}^{<}(t=0) = -i \int^{\infty}_{-\infty} \frac{d\epsilon}{2\pi} {\bf G}^{<}(\epsilon) .
\end{equation}

In the asymptotic limit of large bias difference $(\mu_L \to
\infty, \mu_R \to -\infty)$ the Fermi distribution reduces to $f_{L}(\epsilon) = 1$ and $f_{R}(\epsilon) = 0$. Further considering symmetric coupling and the wide-band limit i.e., $ {\bf \Gamma} (\epsilon) =  {\bf \Gamma} $, the lesser component of the total self-energy (Eq.~\ref{self}) simplifies to,
\begin{eqnarray}
&& {\bf \Sigma}_{L}^{<} (\epsilon)  = i  \Gamma_{L}\begin{pmatrix}1 & 1 \\1 & 1 \end{pmatrix} = i \frac{{\bf \Gamma}}{2} \quad\text{and}\nonumber \\ &&
{\bf \Sigma}_{R}^{<} (\epsilon) =0.
\end{eqnarray}
The Keldysh equation, given in Eq.~\ref{keldysh}, then reduces to
\begin{equation}
{\bf G}^{<}(\epsilon)= \frac{i}{2}  {\bf G}^{r}(\epsilon) {\bf \Gamma} {\bf G}^{a}(\epsilon).
\end{equation}
One further receives, in this limit, the following relation using Eq.~\ref{retarded-advanced}
\begin{equation}
{\bf A}(\epsilon) = i \big({\bf G}^{r}(\epsilon)-{\bf G}^{a}(\epsilon)\big)= {\bf G}^{r}(\epsilon)\, {\bf \Gamma}\, {\bf G}^{a}(\epsilon).
\end{equation}
We can thus write
\begin{equation}
-i {\bf G}^{<}(t=0) = \frac{1}{2} \int^{\infty}_{-\infty}\frac{d\epsilon}{2\pi}{\bf A}(\epsilon) ,
\end{equation}
The above integration over $A_{ii}$ yields $1$
and over $A_{ij}$ yields $0$ for the non-degenerate case whereas for
the degenerate case the integration
over all the components gives the value $\frac{1}{2}$. Therefore, in the asymptotic limit, for the non-degenerate case
\begin{eqnarray} 
\langle d_i^{\dagger} d_i \rangle = \frac{1}{2}, \langle d_{i}^{\dagger} d_{j}\rangle=0
\end{eqnarray} 
and for the
degenerate case, 
\begin{eqnarray}
\langle d_i^{\dagger} d_i \rangle  =\frac{1}{4}, \langle d_{i}^{\dagger} d_{j}\rangle=\frac{1}{4}.
\end{eqnarray}

{\it Charge current:} The above analysis helps us to obtain the asymptotic expression for the steady state charge current for any interacting junction, given as
\begin{eqnarray}
I = \frac{1}{4} \int_{-\infty}^{\infty} d\epsilon {\rm Tr}\Big[{\bf \Gamma} {\bf A(\epsilon)}\Big] = \frac{ \Gamma}{2}.
\end{eqnarray} 
both for degenerate and non-degenerate cases. This yields $I = 1/2$ for $\Gamma = 1$ which exactly matches with our numerical results as shown in Figs.~((\ref{curr}a) and (\ref{ent2}a))\\

{\it Mutual Information and Concurrence:} We next analyze the asymptotic values for the mutual information and the concurrence. The reduced density matrix for the double-dot system is given as in Eq.~\ref{eq32}. Using this, the reduced
density matrix for the the subsystem, where dot $1$ is considered as one
subsystem while the other dot as second subsystem, can be written as
\begin{equation}
\rho_1 = \text{Tr}_2(\rho_{12}), \qquad \rho_2 = \text{Tr}_1(\rho_{12}).
\end{equation}
Calculating these traces we can write the reduced density matrices as 
\begin{equation}
\rho_{1} =\begin{bmatrix}
&u + w_1 &0\\
&0 & v + w_{2}\\
\end{bmatrix}, \ 
\rho_{2} =\begin{bmatrix}
&u + w_2 &0\\
&0 & v + w_{1}\\
\end{bmatrix}.
\end{equation}
The asymptotic values of different correlators in the non-degenerate case are (as calculated above)
\begin{eqnarray}
&\langle n_{1} \rangle = \frac{1}{2},\langle n_{2}\rangle = \frac{1}{2}, z=0, \nonumber \\
&u=v=w_1=w_2=\frac{1}{4},
\label{con1}
\end{eqnarray}
and for the degenerate case 
\begin{eqnarray}
&\langle n_{1} \rangle = \frac{1}{4},\langle n_{2}\rangle = \frac{1}{4}, z=\frac{1}{4},\nonumber \\
& u=\frac{1}{2}, v=0, w_1=w_2=\frac{1}{4}.
 \label{con2}
\end{eqnarray}
The asymptotic value of the mutual information between the two dots can then be calculated using Eq.~\ref{eq23}, which vanishes for the non-degenerate case in this large bias limit, as also reflected in our numerics, Fig.~\ref{stadys2}b. In contrast, for the degenerate case the asymptotic value of mutual information is found to be $\approx 0.43145$, as also obtained numerically in Fig.~\ref{stadys1}b. In a similar manner, the asymptotic value of the mutual information between the bath and dot system can be computed. In the same large bias limit $(\mu_L \to \infty,
\mu_R \to -\infty)$, and given that the
initial dot occupancies are zero, i.e., $n_1 = n_2 = 0$, (see Eq.~\ref{initialcond}), the
overall system  initially is in a pure state and therefore $S_{AB}=0$ and 
$S_{A}=S_{B}$, which imply that the mutual information between the bath and the system is simply related by $S =2S_{A}$. Now using Eq.~\ref{eq32}, in the
non-degenerate scenario one receives the asymptotic value as 
$4 \,\text{ln}2$, whereas in the degenerate case it comes out to be
$2 \, \text{ln}2$ and matches exactly with Fig.~\ref{stadys3}.

The calculation of concurrence using Eq.~\ref{eq33} and Eqs.~(\ref{con1},\ref{con2}) is also straightforward. In the non-degenerate case since $|z|=0$, the concurrence vanishes whereas for the degenerate case it reaches an asymptotic value $1/2$ and further validates our numerical results as shown in Fig.~\ref{stadys1}c and Fig.~\ref{stadys2}c.
\begin{figure*}[t]
\includegraphics[scale=0.75]{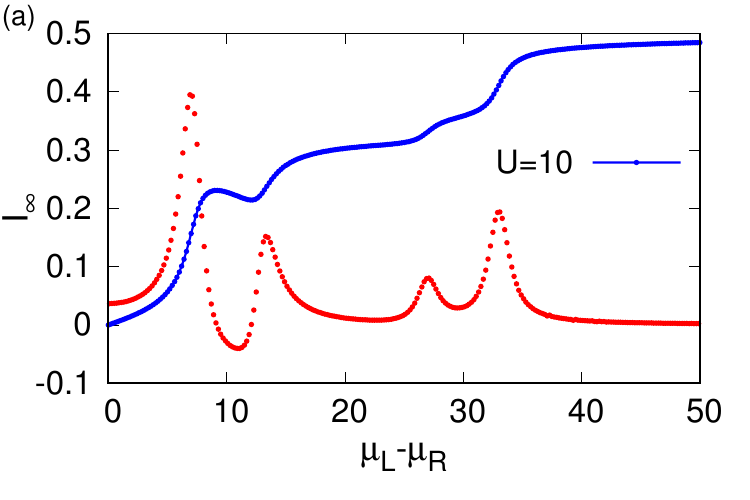}
\includegraphics[scale=0.75]{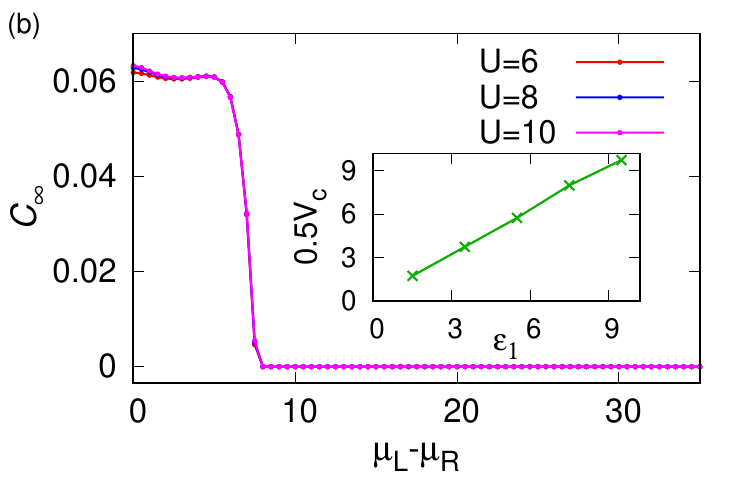}
\includegraphics[scale=0.75]{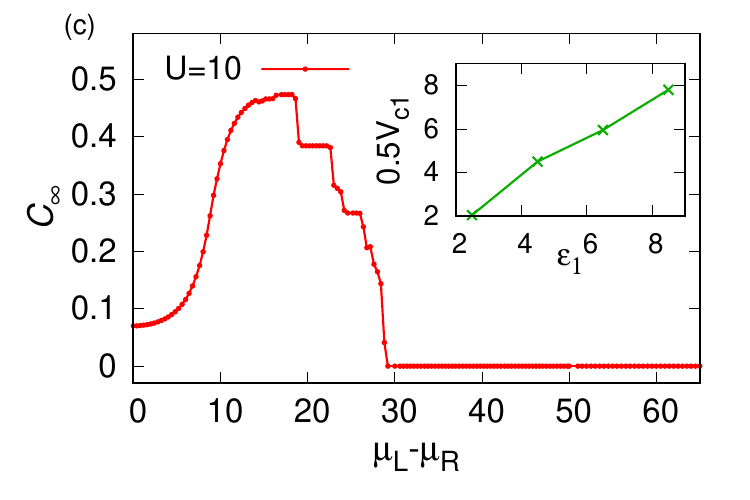}
\caption{(Color online) Steady state properties for the interacting case: (a) Current and (b) concurrence for different dot energies $(\epsilon_{1} = 3.5, \epsilon_{2} = 6.5)$. In (a) the  dotted plot refers to differential conductance $\frac{dI_{\infty}}{dV}$ with $V= \mu_L-\mu_R$ with 5 times enhanced values. The inset in (b) provides the critical voltage $(V_{c} = 2\epsilon_1)$ where the concurrence drops to zero as a function of $\epsilon_1$. (c) Concurrence for same dot energies $(\epsilon_{1} = \epsilon_{2} = 4.5)$. The inset provides the corresponding critical voltage $(V_{c1}=V_c -2U)$ as a function of the dot energy. Here $U=10, k_{B}T=0.05$.  }
\label{int}
\end{figure*}
\begin{figure}[t]
\includegraphics[scale=1.00]{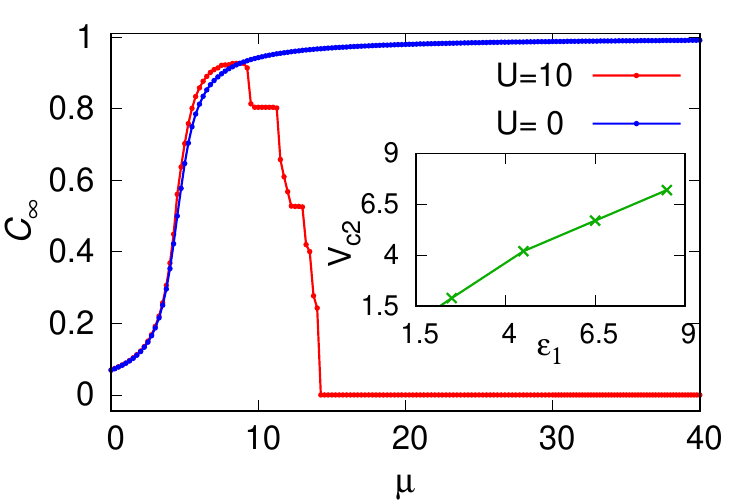}
\caption{(Color online) Steady state plot for the zero bias $(\mu_L = \mu_R)$ concurrence for degenerate dots with (solid, red) and without (solid, blue) interaction. The inset gives the critical value of the bias voltage $(V_{c2}= \frac{V_c}{2}-U)$ as a function of dot energy. Here $ k_{B}T=0.05,\epsilon_1 = \epsilon_2= 4.5 $ and other parameters are taken same as before. }
\label{int1}
\end{figure}
%%%%%%%%%%%%%%%%%%%%%%%%%%%%%%%%%%%%%%%%%%%%%%%%%%%%%%%%
%%%%%%%%%%%%%%%%%%%%%%%%%%%%%%%%%%%%%%%%%%%%%%%%%%%%%%%%
\section{Interacting Dots ($U \neq 0$)} 
\subsection{Formalism}
In this section we extend our study for the interacting case. The Hamiltonian for the spinless double quantum dot with inter-site Coulomb  interaction is given in Eq.\ref{eq1}. The NEGF calculation for the interacting model follows along the similar lines as the non-interacting case, however, in this case higher order Green's functions needs to be computed and a suitable approximation scheme is therefore required to achieve closure in the EOM approach. 
%Closure needs to be imposed with the help of a suitable  approximation scheme.\\

Let us first start writing the EOM for the interacting Green's function for the dots, 
\begin{eqnarray}
\Big[i \frac{\partial}{\partial t}-\epsilon _{i} \Big]{G}_{{i}{j}}(t,t')  =  \delta_{ij}  \delta  (t&&-t')    +  \sum _{k,\alpha =L,R} (t^i_{k,\alpha })^* G_{k{j}}(t,t') \nonumber \\ && + U G_{{i}{m}{j}}(t,t').  \hspace{0.45cm}  m\neq i
\end{eqnarray}
Where due to interaction higher order (two-particle) Green's function appears, such as,
\begin{eqnarray}\label{eq:Gimj}
G_{{i}{m}{j}}(t, t') =  -i \langle T d_{i}(t)n_{m}(t)d_{j}^{\dagger}(t')\rangle, \hspace{0.35cm}   m\neq i
\end{eqnarray}
 and we need to compute the corresponding EOM as well,
\begin{eqnarray}
\Big[i \frac{\partial}{\partial t}-\epsilon _{i}-U \Big]{G}_{{i}{m}{j}}(t,t') =  && \delta_{ij}\delta(t-t')\langle n_{m}(t)\rangle  \\ && + \sum _{k\alpha = R, L} (t^i_{k,\alpha })^* G_{k{m}{j}}(t,t').\nonumber
\end{eqnarray}
At this point, let us introduce the Green's function associated only with the dots as,
\begin{equation}
\Big[ i\frac{\partial}{\partial t} - \epsilon_{i} -U \Big] g_{ij U}(t,t') = \delta_{ij}\delta(t-t').
\end{equation}
In the above EOM we use the steady state property, i.e., $\dot{n}_{m}(t) = 0$ and $\langle n_{m}(t)\rangle$ is the average dot occupancy.
This procedure yields a higher order correlator, mixing the dots and the leads
\begin{eqnarray}
G_{k{m}{j}}(t, t') =  -i \langle T c_{k \alpha}(t)n_{m}(t)d_{j}^{\dagger}(t')\rangle. 
\end{eqnarray}
So, a further EOM is necessary: 
\begin{equation}\label{eq:Gkmj}
\Big[i \frac{\partial}{\partial t}-\epsilon _{k\alpha} \Big]{G}_{k{m}{j}}(t,t')  = \sum_{l=1,2} t_{k \alpha }^{l} G_{{l}{m}{j}}(t,t')
\end{equation} 
In the above equation the Green's functions (in the right hand side of
the equation) for which $``l=m"$ vanish because multiple occupancy on
a dot at the same time is forbidden due to Pauli exclusion
principle. The mixed correlators containing operators
  from the lead and the dots such as $\langle c_{k\alpha}
  d_{i}^{^{\dagger}}\rangle$ are set to zero thus treating the coupling
  of the leads and system upto second order in $t_{k\alpha}$. Also
  the Green's functions like Eq.~\ref{eq:Gimj} are decoupled using the
  mean field approach as $G_{{i}{m}{j}}(t, t')= \langle n_{m}\rangle
  G_{ij}(t,t')$. These two approximations have been shown to yield reliable results
  for a sufficiently high temperature~\citep{meir1991transport,sun2002double,levy2013steady}.
This approximation leads to a closure for the cascade of Green functions in the EOM
approach. As done for the non-interacting case, one can then write
down an EOM for the contour-ordered version and obtain all the
components in real time using Langreth's thorem. The various
correlators involving the system operators can then be computed self
consistently using
\begin{eqnarray}
\langle d_{{i}}^{\dagger}d_{j}\rangle = -i \int_{-\infty}^{\infty} \frac{ d\epsilon}{2 \pi} G^<_{{ji}}(\epsilon).
\end{eqnarray}
%=======================================================================================================================
\subsection{Results}
To understand the effect of Coulomb interaction on entanglement properties we first check the transport properties by computing the current and the corresponding differential conductance.
We follow the NEGF method and use Eq.~\ref{eq29} to evaluate the current.
In Fig.~\ref{int}(a)) the standard Coulomb blockade effect is evident and the corresponding peaks are reflected in the differential conductance. In Particular, the blockade
starts when the bias voltage corresponds to
$2\epsilon_1,\ 2\epsilon_2,\ 2(\epsilon_1+U)$ and $2(\epsilon_2
+U)$.  Note that, once again the asymptotic value for the current is $1/2$.

Next, to compute the concurrence following Eq.~\ref{eq33} one needs to calculate different two-point and four-point correlators.   Since for the interacting case correlators like $\langle n_in_j\rangle$ can not be computed following the same trick as done in the non-interacting case, we employ here the Hartree-Fock approximation scheme ~\citep{dhar2006nonequilibrium,agarwal2006conductance} i.e.,
\begin{equation}
n_in_j = \frac{1}{2} \Big[ \langle d_j^{\dagger}d_j\rangle d_i^{\dagger}d_i + d_j^{\dagger}d_j \langle d_i^{\dagger}d_i\rangle -  \langle d_i^{\dagger}d_j\rangle d_j^{\dagger}d_i-\langle d_j^{\dagger}d_i\rangle d_i^{\dagger}d_j \Big].
\end{equation}
Fig.~(\ref{int}b) and (\ref{int}c) display the steady state concurrence for non-degenerate and degenerate case in presence of the interaction. For the non-degenerate case, a similar argument as the non-interacting model is admissible. For low bias, the concurrence is small and drops sharply to zero for $\mu_L-\mu_R > 2 \epsilon_1$. As evident from the numerics, the interaction has practically no effect on concurrence, in this case. 

In contrast, in the degenerate scenario with finite interaction $U$
(Fig.~\ref{int}c), the behaviour of concurrence is rich and can be
analyzed in three distinct regimes for the bias voltage: (i) when the
bias voltage is below the energy level of the dot i.e., $ \mu_L-\mu_R
\leq 2 \epsilon$, the concurrence increases monotonically just like in
the non-interacting case. (ii) For bias voltage in the range $2
\epsilon \leq \mu_L-\mu_R \leq 2(\epsilon+U)$, the dot levels are
non-degenerate due to the finite interaction $U$ and while one dot is
occupied by electron the other electron needs to have an effective
$2(\epsilon+U)$ energy to tunnel. Even if electron manages to tunnel
through this level it leads to a separable state and entanglement
between the two dots starts to decrease.  (iii) further increase in
the bias i.e., $\mu_L-\mu_R > 2(\epsilon + U)$ the concurrence drops
to zero as after filling one dot with energy $\epsilon$, the other
electrons can tunnel through the shifted $2(\epsilon+U)$ level,
leaving the two dots in a separable state.  The inset of
Fig.~(\ref{int}c) shows the plot for the critical bias, at which
concurrence hits zero, with the energy level of the dot $\epsilon$ and
a linear dependence is observed. In this figure
  the concurrence shows a step-like fall with the bias. 
  This is due to the uneven change in the average electron number $\langle n_{i}\rangle$
  with bias at higher values of bias, which in turn is a numerical
  artifact of the self consistently approximate computation of the average
  electron number.

In order to contrast transport and quantum correlations, we further investigate the situation when the two fermionic baths are maintained at the same chemical potential ($\mu_L =\mu_R =\mu$). Due to zero bias difference, net current into and out of the dots cancel out providing a dynamical zero, independent of the chemical potentials of the leads and the dot energy levels. However, under the same scenario, the entanglement between the dots depends on the value of the chemical potential, as well as the dot energies. In Fig.~(\ref{int1}) we  display the concurrence starting with the degenerate dot level case $(U=0)$. A high value for the concurrence is observed with increasing chemical potential in the  non-interacting case and finally satures to the value 1.0 representing  maximally entangled state for the dots. However, turning on the interaction lifts the degeneracy and for large chemical potential the two dots achieve a separable state leading to a zero concurrence. The inset in Fig.~(\ref{int1}) shows the linear dependence of the chemical potential at which concurrence reaches zero with the dot energy level. Again the step-like fall in concurrence here is caused by the numerical artifacts in the self consistent calculation of average electron number.
\section{Summary}
In summary, we have studied quantum transport and entanglement properties for out-of-equilibrium spinless parallel interacting quantum double
dot setup.  Employing NEGF and exact numerical approaches we have investigated quantum dynamics and the steady state properties for non-equilibrium charge current, mutual information and concurrence.
It is found that both the transient and steady state behaviour of these observables is critically dependent on whether or not the dot energies are degenerate. In addition, strong correlations between these observables is found in both transient and steady state regimes. For example, In the non-degenerate case, for high bias, both the mutual information and concurrence approaches to zero in the steady state. Whereas for the degenerate case, a high value for both these observables is found to exist. In contrast, the asymptotic value for the current remain insensitive both in degenerate and non-degenerate limit. The appearance of plateau in the current for the non-degenerate case also reflected in mutual information between the dots as well as between the dots and the bath. 
 
For the interacting case, we employ the Hartree-Fock approximation scheme to compute the steady state concurrence. 
The effects of interaction are once
again tied up with degeneracy effects. The non-degenerate case is
largely independent of interactions whereas in the degenerate case, the concurrence increases for small bias but drops to zero beyond a critical bias because of the lifting of the degeneracy. The characteristic value of this bias though, is dependent on the strength of interaction unlike in the non-degenerate case.

The future work will direct towards understanding the entanglement and transport properties for extended many-body quantum systems, in particular, scaling properties with the system size.

\section*{Acknowledgements}
AD thanks Sreeraj Nair for helpful discussions. BKA thanks IISER-Pune
for the start up funding. BKA and AS jointly acknowledge the
International Centre for Theoretical Sciences for facilitating discussions during a visit for participating in the program -  Indian Statistical Physics Community Meeting 2018 (Code: ICTS/Prog - ISPCM2018/02). 
AS is grateful to SERB which funds AD's JRF via the startup grant (File Number: YSS/2015/001696).
D.S.B acknowledges PhD fellowship support from UGC India.
\bibliography{ref}
\end{document}